\documentclass{esannV2}
\usepackage[latin1]{inputenc}
\usepackage{amssymb,amsmath,array}
\usepackage{hyperref}
\usepackage{multirow}
\usepackage{graphicx,graphics}
\usepackage{subfig}
\usepackage{booktabs}  
\usepackage{algorithm,float}
\usepackage[noend]{algpseudocode}
\usepackage{bm}

\newcommand{\vect}{\mathbf{t}}
\newcommand{\vecy}{\mathbf{y}}

\begin{document}
\title{MultiMBNN: Matched and Balanced Causal Inference with Neural Networks}

\author{Ankit Sharma$^1$, Garima Gupta$^1$, Ranjitha Prasad$^2$, Arnab Chatterjee$^1$, \\Lovekesh Vig$^1$ and Gautam Shroff$^1$
%
%
\vspace{.3cm}\\
%
1- TCS Research, Delhi, India 
%
\vspace{.1cm}\\
2- Indraprastha Institute of Information Technology, Delhi, India \\
}


\maketitle
\begin{abstract}
Causal inference (CI) in  observational studies has received a lot of attention in healthcare, education, ad attribution, policy evaluation, etc. Confounding is a typical hazard, where the context affects both, the treatment assignment and response. In a multiple treatment scenario, we propose the neural network based \texttt{MultiMBNN}, where we overcome confounding by employing generalized propensity score based matching, and learning balanced representations. We benchmark the performance on synthetic and real-world datasets using PEHE, and mean absolute percentage error over ATE as metrics. \texttt{MultiMBNN} outperforms the state-of-the-art algorithms for CI such as \texttt{TARNet} and Perfect Match (\texttt{PM}).
\end{abstract}

\section{Introduction}
The primary goal in causal inference (CI) is to uncover the cause-effect relationship between entities, often construed as a problem of quantifying the effect of treatments on individuals. Randomised control trials, the most popular choice for obtaining causal relationships, are expensive and entail several logistical and ethical constraints. Causality is a crucial paradigm in several domains where observational data is available, such as healthcare~\cite{stuart2010matching}, socioeconomic studies~\cite{athey2015machine}, advertising~\cite{bottou2013counterfactual}, etc. An impediment to  CI in observational studies is the presence of \emph{confounding}, where assignment and the  response to the treatment depends on context covariates, resulting in selection bias. 

In order to abate the effect of confounding, the discrepancy between the distribution of individuals receiving different treatments is minimized, in order to emulate a randomized trial. In the binary treatment case, confounding is addressed using statistical approaches such as sub-classification, weighting, imputations, and propensity score (PS) matching for unbiased per-individual causal estimates. Deep neural networks~(DNN) based techniques propose counterfactual distribution modelling~\cite{louizos2017causal}, and learning balancing representation to minimize selection bias~\cite{johansson2016learning, shalit2017estimating, sharma2019metaci}.
In literature, the multiple treatments scenario is interpreted as different dosage levels of a single treatment, or being one of several treatments. For the latter case, matching and sub-classification techniques have been proposed~\cite{stuart2010matching}, and in particular, generalized propensity score (GPS) based matching was proposed~\cite{imai2004causal} as an accurate metric in the multiple treatment scenario~\cite{stuart2010matching}. DNN-based techniques for CI include counterfactual distribution learning~\cite{yoon2018ganite}, Gaussian process based modeling\cite{alaa2017bayesian}, and  PS matching based Perfect match~\cite{schwab2018perfect}. Here, we propose a novel DNN for counterfactual inference, which overcomes confounding by  leveraging on both, GPS based matching and learning balancing representations.

\noindent \textbf{Contributions:} We propose a novel framework for counterfactual inference in the presence of confounding due to multiple treatments. We assume strong ignorability and no hidden confounding \cite{johansson2016learning, shalit2017estimating, sharma2019metaci}. We optimize DNN models via mini-batch stochastic gradient descent (SGD) to predict both, the factual and counterfactual response of a given individual to any one of the treatments. We propose the GPS based matching, along with learning balanced representations to address confounding. We estimate the GPS by training a predictive model and we use this GPS to \emph{match} every sample within a minibatch with its nearest neighbours. In order to learn the balanced representation, we propose a loss function using the pairwise minimum mean discrepancy (MMD) metric. 
The novelty of this work is two-fold. First, we show that GPS based matching leads to more accurate counterfactual inference as compared to~\cite{schwab2018perfect}. Next, we generalize the balancing representation based loss function~\cite{johansson2016learning, shalit2017estimating} to multiple treatment scenario. On synthetic and real-world datasets, we demonstrate that by combining matching and the generalized loss function, we outperform the state-of-the-art CI techniques such as perfect match (PM) \cite{schwab2018perfect} and TARNet \cite{shalit2017estimating}. We use precision in estimation of heterogeneous effect ($PEHE$) and mean absolute percentage error($MAPE$) over the average treatment effect ($ATE$) as metrics.  

\vspace{-2mm}
\section{\texttt{MultiMBNN} Causal Inference Model}

In this section, we describe the preliminaries of the CI framework in the multiple treatment scenario, followed by details of the proposed \texttt{MultiMBNN} framework. 
\vspace{-2mm}
\subsection{Causal Inference Preliminaries}
\vspace{-2mm}
We consider observational training data $D_{CI}$, comprising of $N$ samples, where each sample is given by $\{x_i,\vect,\vecy_{i}\}$. Each individual (also called \textit{context}) $i$ is represented using covariates given by $x_i$ for $1\leq i \leq N$. An individual is subjected to one of the $K$ treatments given by $\vect = [t_1, \hdots, t_K]$, where each entry of $\vect$ is binary, i.e., $t_k \in \{0,1\}$. Here, $t_k = 1$ implies that the $k$-th treatment is given. We assume that only one treatment is provided to an individual $i$ at any given point in time, and hence, $\vect$ is an one-hot vector. Accordingly, the response vector for the $i$-th individual is given by $\vecy_i \in \mathbb{R}^{K \times 1}$, i.e., the outcome is a continuous random vector with $K$ entries denoted by $y_{ik}$, the response of the $i$-th individual to the $k$-th treatment. We define counterfactual as the $K-1$ alternate treatments which are unobserved for an individual.

Our goal is to train a DNN model to overcome confounding and perform counterfactual regression, i.e., to predict the response given any context and treatment. We address the issue of confounding using both, matching and learning balanced representations. In the sequel, we describe the matching method used and the loss function that caters to a multiple treatment scenario. 

\vspace{-2mm}
\subsection{Generalized Propensity Score Matching}
\label{sec:GPS}
Propensity score (PS) based matching is a well-known technique used to induce the effect of randomized experiment by obtaining  similar covariate distributions across treated populations~\cite{stuart2010matching}. Here we employ matching based on Generalised Propensity Score (GPS), which is a more relevant score for the  multiple treatment scenario. 
PS is the conditional probability of a given individual $x_i$ receiving a treatment $t_k$, i.e., $p(t_k|x_i)$. Accordingly, the  GPS vector is defined as $p(\vect|x_i) = [p(t_1|x_i),p(t_2|x_i),\hdots,p(t_K|x_i)]$. In practice, we do not have access to the GPS vector, and hence, we estimate it by training a predictive model. In this work, we train an SVM or random forest . We use this tuned model to predict PS, $p(\vect|x_i)$, on the training data used for causal inference. In order to avoid overfitting, we first obtain $l$ nearest neighbors, and pick one out of these $l$ samples at random for each counterfactual treatment of $x_{i}$.
We employ the GPS vector $p(\vect|x_i)$ for batch augmentation in every minibatch of SGD. For every sample within a minibatch, $K-1$ closest neighbour samples are obtained. For instance, consider a sample $x_i$ and its factual treatment $t_i$. We propose the GPS-based matching strategy $M_{GPS}$, which selects a neighbour $x_j$ with observed treatment $t_j$ such that $t_j \neq t_i$ and $d_{GPS}(i,j)$ is minimum. Here, $d_{GPS_{i,j}}$ is defined as $d_{GPS}(i,j) = \sum_{k=1}^{K}|p(t_k|x_i)-p(t_k|x_j)|$.
On the other hand, the PS matching strategy $M_{PS}$ \cite{schwab2018perfect} selects a neighbour $x_j$ with observed treatment $t_j$ such that $t_j \neq t_i$ and $d_{PS}(i,j)$ is minimum where $d_{PS}(i,j) = |p(t_j|x_i)-p(t_j|x_j)|$. Albeit its popularity, PS based matching has been described as inadequate and it sometimes leads to imbalance in parametric models due to model dependence~\cite{king2019propensity}. Hence, we learn balancing representations to achieve better performance.

\subsection{Learning Balancing Representations}
\label{sec:balancing}
In addition to overcoming the imbalance using matching, we also propose learning balanced representation using DNN. In~\cite{johansson2016learning}, the authors perform counterfactual inference by generalizing the factual to counterfactual distribution, for the binary treatment scenario.  We extend this framework from binary to a multiple treatment scenario, and modify the loss function as follows:
\begin{equation}
\mathcal{L}\left(\alpha, \gamma\right) = \tfrac{1}{N} \sum_{i=1}^{N}{\left(h\left(\Phi(x_i),t_k\right) - y_{ik}\right)}^2 + \alpha \sum_{m = 1}^K\sum_{q = 1}^{m-1} \mbox{disc}\left(\hat{p}^{m}_{\Phi},\hat{p}^{q}_{\Phi}\right) + \gamma \mathcal{R}(h),
    \label{eq:loss}
\end{equation}
where $\alpha, \gamma > 0$ are hyperparameters that control the strength of the imbalance penalties, $\mathcal{R}(h)$ is a model complexity term, $\hat{p}^{\textnormal{m}}_{(\cdot)}$ and $\hat{p}^{\textnormal{q}}_{(\cdot)}$ represent the distribution w.r.t. the $m$-th treatment and the $q$-th treatment, respectively, and $\mbox{disc}(\cdot,\cdot)$ is the minimum mean discrepancy measure as defined in~\cite{shalit2017estimating}. We learn the balancing representation $\Phi(\cdot)$ and the hypothesis $h(\cdot)$ jointly by training a deep neural network using a loss function that incorporates the factual and the imbalance error as depicted in Fig.~\ref{fig:PropMultiMBNN}. In \eqref{eq:loss}, the first term on the right hand side represents the factual loss. The second term computes the pairwise minimum mean discrepancy between factual distributions of different treatments. The loss function in \eqref{eq:loss} is a generalisation of \cite{shalit2017estimating} to the multiple treatment scenario, it reduces to the one proposed in~\cite{shalit2017estimating} for $K = 2$.

\begin{algorithm}
\caption{\texttt{MultiMBNN} algorithm}\label{alg:MultiMBNN}
\begin{algorithmic}[1]
\Procedure{MultiMBNN}{D}
\State Split dataset $D$ into \textit{$D_{CI}$} for CI, and \textit{$D_{PS}$} to compute GPS.
\State Divide $D_{CI}$ into train ($D_{CI,t}$), validation and test sets.
\State Obtain GPS $p(\vect|x_i), \forall i$ in $D_{CI,t}$, as described in Sec.~\ref{sec:GPS}.
\State Divide $D_{CI,t}$ into batches with each batch being $M_{B}$
\For{$E$ epochs and $M_B \in D_{CI,t}$}
\State $\tilde{M}_B \gets$ augment $M_{B}$ using $M_{GPS}$, as described in Sec.~\ref{sec:GPS}
\State Update $\Phi(\cdot)$, $h(\cdot)$ using input $\tilde{M}_B$ by minimizing Eq.~\ref{eq:loss}
\EndFor
\Return $\Phi(\cdot)$, $h(\cdot)$ 
\EndProcedure
\end{algorithmic}
\end{algorithm}
\vspace{-4mm}
\subsection{Proposed Approach: \texttt{MultiMBNN}}

\begin{figure}[h]
    \centering
    \includegraphics[width=9cm]{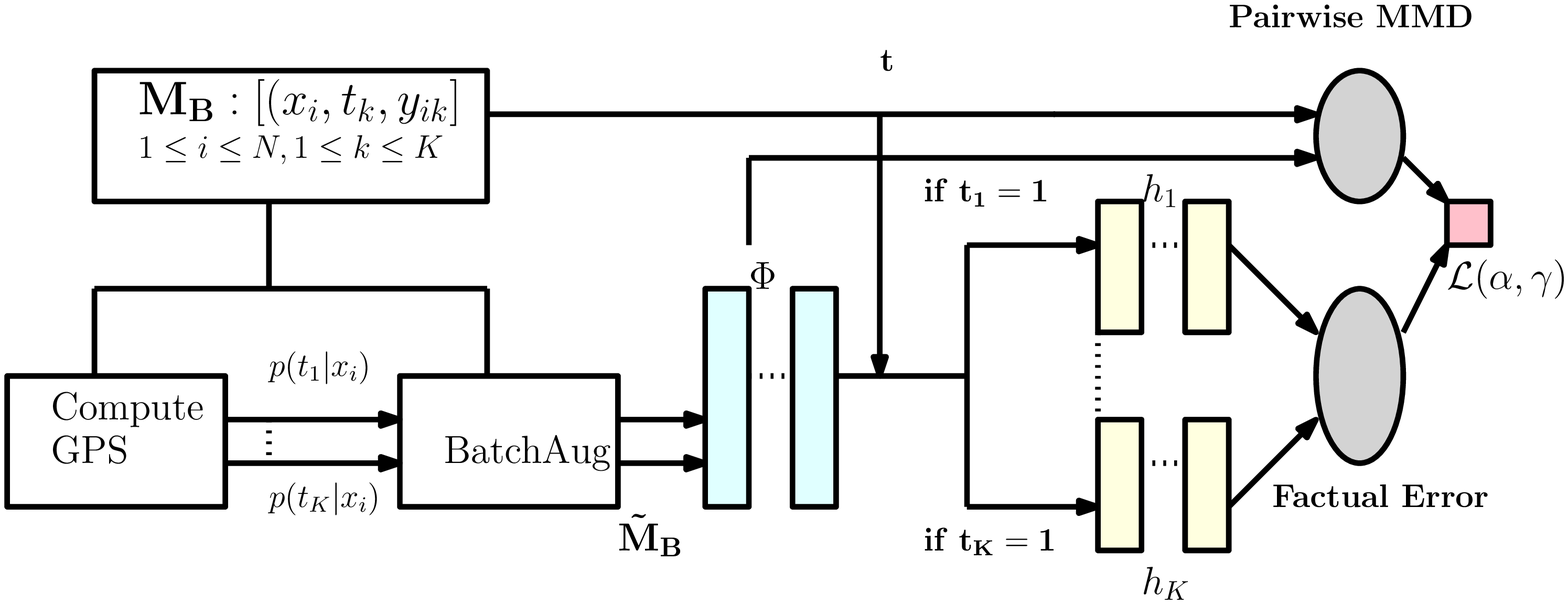} 
    \caption{\texttt{MultBNN}: Proposed Neural network architecture}%
    \label{fig:PropMultiMBNN}
\end{figure}

\noindent We propose the \texttt{MultiMBNN} algorithm as described in Algorithm~\ref{alg:MultiMBNN}. As discussed in the previous subsections, we perform batch augmentation based on GPS, and train a DNN to learn the balancing representation, $\Phi$ and the hypothesis layers $h_1, \hdots, h_K$, one for each treatment, using the augmented minibatches, via SGD based training. The proposed neural network architecture is depicted in Fig.~\ref{fig:PropMultiMBNN}.
\vspace{-4mm}
\section{Experiments and Results}

\begin{figure}[h!]
\label{fig:PlotCFEKappa}
\centering
\hspace{-0.4cm}
\includegraphics[scale=0.250]{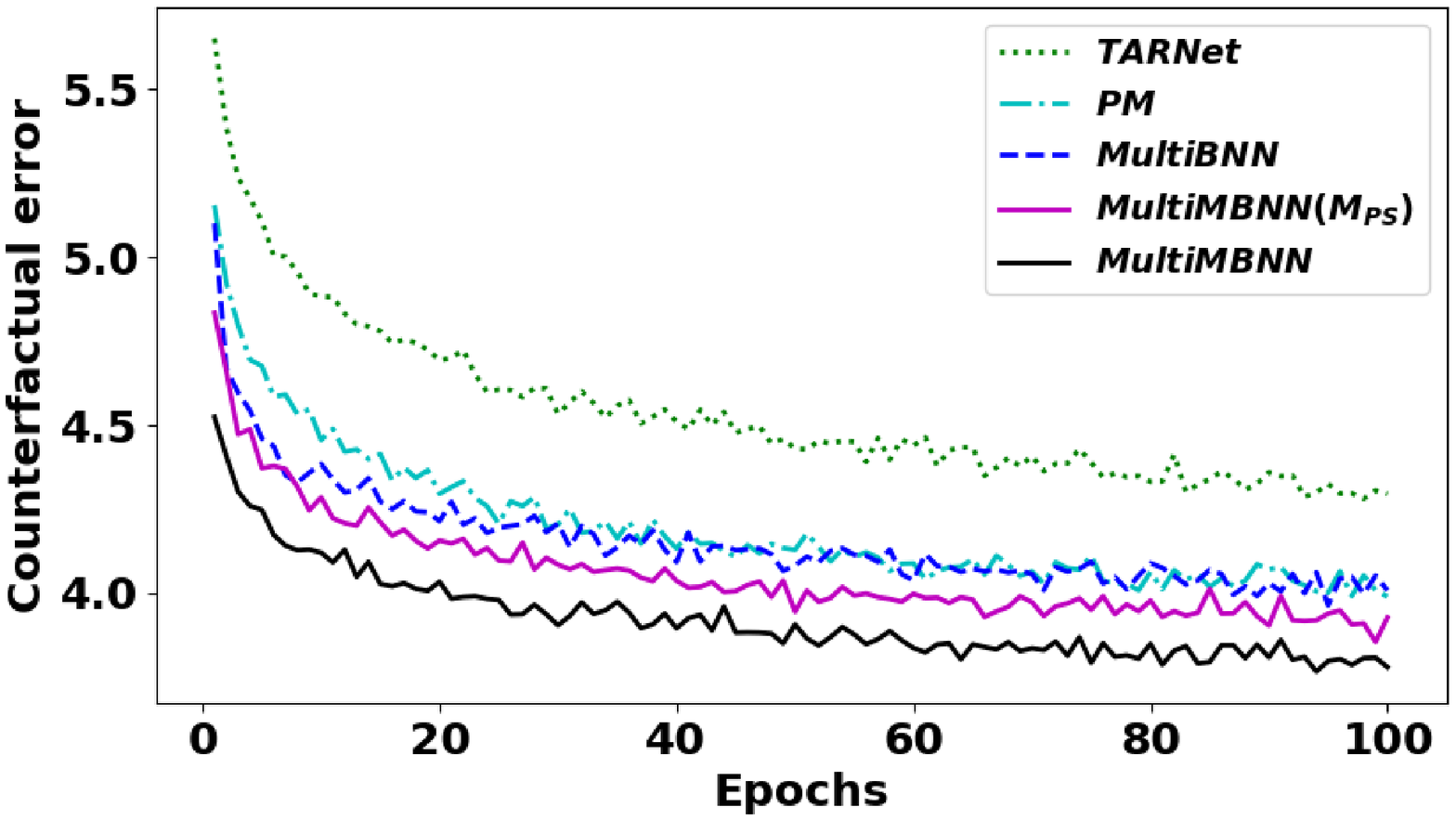}
\includegraphics[scale=0.250]{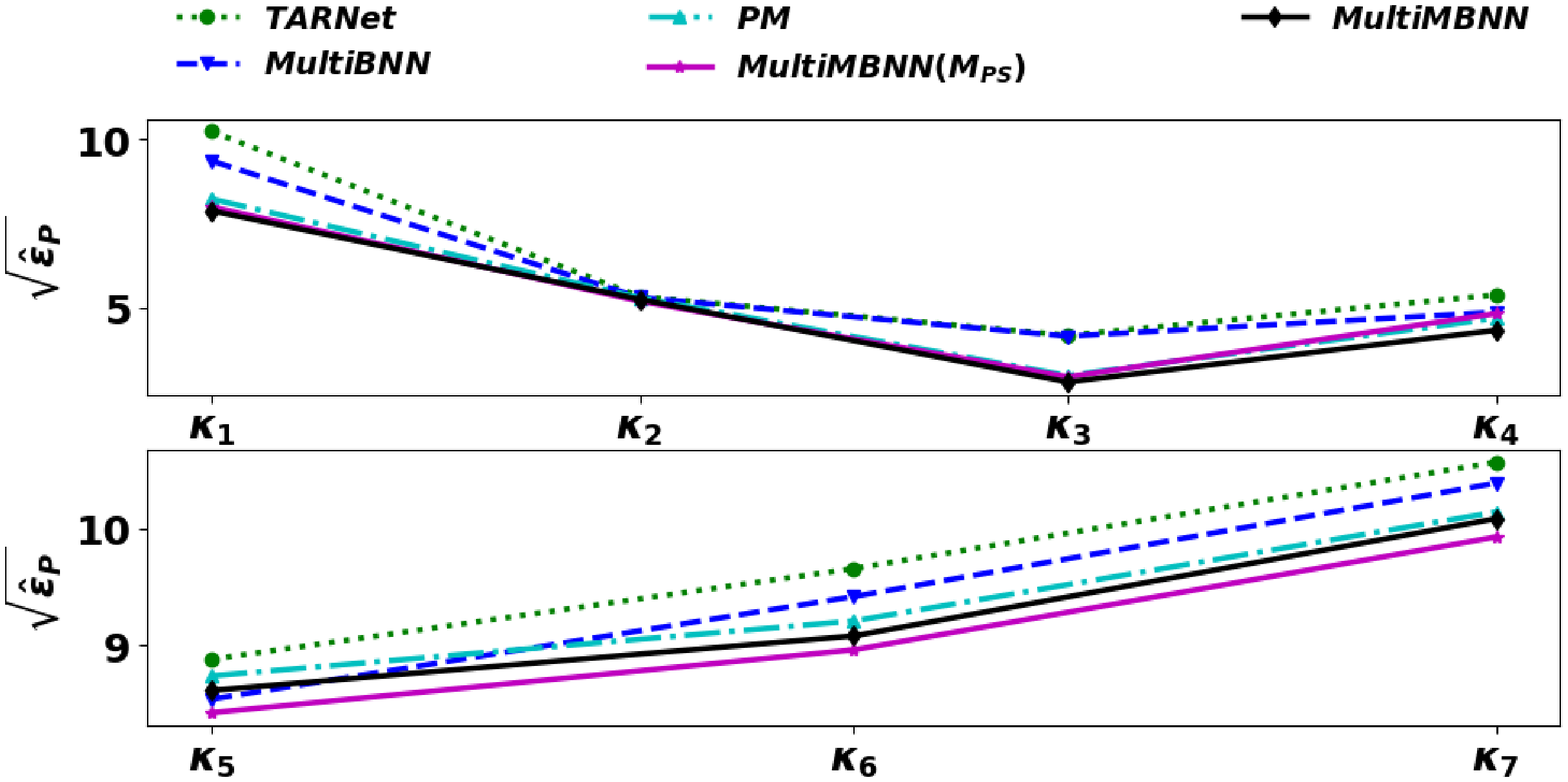}
\caption{Comparison of CI frameworks: Counterfactual error across epochs (Syn4, $\kappa_1$) on the left, and $\sqrt{\hat\epsilon_{P}}$ vs. $\kappa$ on the right (\textit{top right}: Syn4 with $\kappa_{1} < \kappa_2 < \kappa_3 < \kappa_4$, \textit{bottom right}: NEWS4 with $\kappa_5 < \kappa_6 < \kappa_7$).}\label{Fig:PlotCFEKappa}
\end{figure}

We illustrate the performance of \texttt{MultiMBNN} algorithm on synthetic (named as Syn) \cite{sun2015causal}, semi-synthetic NEWS \cite{schwab2018perfect} and cancer genome TCGA\footnote{https://github.com/d909b/perfect\_match} datasets. We obtain $N = 15000$ samples with $10$ covariates for the synthetic dataset with. Here, $\kappa$ accounts for the treatment assignment bias. We employ the DGP in~\cite{schwab2018perfect} and $5772$ bag-of-words context covariates to generate $N = 10000$ samples of the NEWS dataset. In the case of synthetic and NEWS datasets, we generate data for $K = 4,6,8$, referred to as 'name of the dataset', followed by $K$. The TCGA dataset consisting of $10000$ samples with $20547$ covariates is obtained using the DGP in~\cite{schwab2018perfect} with $K=4$ (`TCGA4'). We use $PEHE$ (denoted as $\epsilon_P$) as defined in~\cite{schwab2018perfect}, and MAPE over $ATE = \tfrac{1}{N}\sum_{i=1}^{N}|y_{ik} - \tfrac{1}{K-1}\sum_{j=1,t_{j}\neq t_{k}}^{K-1}y_{ij}|$ \cite{sharma2019metaci}.
We baseline the proposed algorithm using  \texttt{TARNet}~\cite{shalit2017estimating}, \texttt{MultiBNN} which learns balanced representations as described in Sec.~\ref{sec:balancing}, \texttt{PM} is as described in~\cite{schwab2018perfect}, and \texttt{MultiMBNN} ($M_{PS}$), which uses $M_{PS}$ (not $M_{GPS}$) along with balanced representation. We demonstrate the performance of \texttt{MultiMBNN} algorithm using several experimental settings. First, we illustrate the effect of treatment assignment bias using the parameter $\kappa$. As illustrated in Fig.~\ref{Fig:PlotCFEKappa}, for Syn4 \texttt{MultiMBNN} performs the best for $\kappa_{3}$ since imbalance amongst treatments leads to one of the four treatments to be suppressed resulting in a large counterfactual error and hence, an elbow point. For NEWS4, $\kappa_{5}$ has the least counterfactual error since imbalance amongst treatment groups is minimum, leading to near uniform distribution of population samples in all groups. 
\begin{figure}[h!]
\centering
\hspace{-0.4cm}
\includegraphics[width=0.99\linewidth]{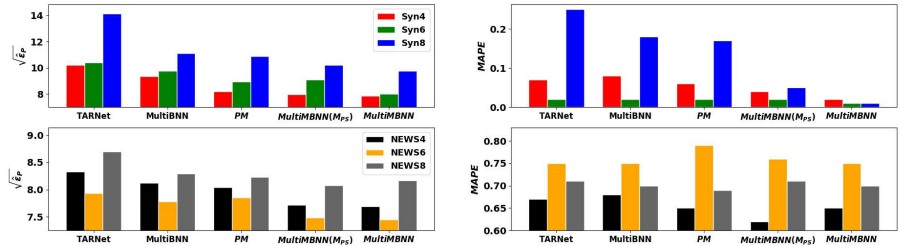}
\caption{$\sqrt{\hat\epsilon_{P}}$ and MAPE for Syn($\kappa_1$) and NEWS($\kappa_6$) datasets for varying $K$.}\label{Fig:PlotVaryingT}
\end{figure}
\begin{table}[h!]
  \centering
   \resizebox{\linewidth}{!}{
  \begin{tabular}{|c|c|c|c|c|c|}
    \hline
    \scriptsize{Metrics,Dataset} & \scriptsize{\texttt{TARNet}} & \scriptsize{\texttt{MultiBNN}} & \scriptsize{\texttt{PM}} & \scriptsize{\texttt{\texttt{MultiMBNN}}($M_{PS}$)} & \scriptsize{\texttt{MultiMBNN}}\\
    \hline
    $\sqrt{\hat\epsilon_{P}}$, Syn4& 10.21 $\pm$ 0.56 & 9.34 $\pm$ 0.61 & 8.21 $\pm$ 0.35 & 7.98 $\pm$ 0.23 & \textbf{7.86 $\pm$ 0.37} \\
    MAPE, Syn4& 0.07 $\pm$ 0.02 & 0.08 $\pm$ 0.03 & 0.06 $\pm$ 0.02 & 0.04 $\pm$ 0.01 & \textbf{0.02 $\pm$ 0.02} \\
    $\sqrt{\hat\epsilon_{P}}$, NEWS4& 9.70 $\pm$ 1.15 & 9.37 $\pm$ 0.90  & 9.16 $\pm$ 0.80  & 8.99 $\pm$ 0.94 & \textbf{8.96 $\pm$ 0.92}  \\
    MAPE, NEWS4& 0.81 $\pm$ 0.11 & 0.81 $\pm$ 0.10  & 0.81 $\pm$ 0.12 & \textbf{0.80 $\pm$ 0.13}  & 0.82 $\pm$ 0.12 \\
    $\sqrt{\hat\epsilon_{P}}$, TCGA4& 29.45 $\pm$ 3.48 & 26.15 $\pm$ 3.29 & 23.57 $\pm$ 1.10 & 23.18 $\pm$ 1.39 & \textbf{21.47 $\pm$ 0.96} \\
    MAPE, TCGA4& 0.93 $\pm$ 0.14 & 0.84 $\pm$ 0.10 & 0.92 $\pm$ 0.07 & \textbf{0.78 $\pm$ 0.03} & 0.80 $\pm$ 0.08 \\
    \hline
  \end{tabular}
  }
  \caption{$\sqrt{\hat\epsilon_{P}}$ and MAPE for multiple runs of Syn4, NEWS4, TCGA4 (fixed $\kappa$). }\label{Table:multipleRuns}
\end{table}

In  Fig.~\ref{Fig:PlotVaryingT}, we illustrate the performance of the proposed algorithms and the baselines with varying $K$ for a fixed $\kappa$. We see that the \texttt{MultiMBNN} algorithm outperforms the baselines by large margins. Further, we simulate \texttt{MultiMBNN} with different initial seed-points maintaining $K$ and $\kappa$ fixed, and report the mean and standard deviation in  $\sqrt{\hat\epsilon_{P}}$ and $MAPE$ for all baselines in Table~\ref{Table:multipleRuns}. We infer that \texttt{MultiMBNN} which incorporates both matching and DNN based balancing, fairs considerably well over all the baselines.
\vspace{-2mm}
\section{Conclusions}
In this work, we propose \texttt{MultiMBNN} algorithm which addresses the inadequacies of the matching framework by learning the balanced representations in multiple treatment causal inference scenario. We demonstrate that \texttt{MultiMBNN} outperforms the state-of-the-art techniques for multiple treatments and also for datasets with thousands of potential co-variate confounders. In future, we shall extend this algorithm for handling sparsity in the presence of large number of treatments.


\begin{footnotesize}

\end{footnotesize}


\end{document}